\documentclass[sigconf,screen,authorversion,nonacm]{acmart}
\AtBeginDocument{%
  }

\setcopyright{acmlicensed}
\copyrightyear{2018}
\acmYear{2018}
\acmDOI{XXXXXXX.XXXXXXX}
\acmConference[Conference acronym 'XX]{Make sure to enter the correct
  conference title from your rights confirmation email}{June 03--05,
  2018}{Woodstock, NY}
\acmISBN{978-1-4503-XXXX-X/2018/06}

\acmSubmissionID{1844}

\usepackage{makecell}
\usepackage{booktabs}
\usepackage{multirow}
\usepackage{array}
\usepackage{xcolor}
\usepackage{graphicx}
\usepackage{enumitem}
\usepackage{hyperref}
\usepackage{cleveref}
\renewcommand\footnotetextcopyrightpermission[1]{}  
\usepackage{booktabs}
\usepackage{multirow}
\usepackage[table]{xcolor}
\definecolor{lightgray}{gray}{0.94}

\begin{document}

\title{Between Safe Boundaries: Exploiting Temporal Consistency for Jailbreaking Text-To-Video Generation Models }

\author{Xingkai Peng}
\email{penglevy@mail.ustc.edu.cn}
\affiliation{%
  \institution{University of Science and Technology of China}
  \city{Hefei}
  \state{Anhui}
  \country{China}
}

\author{Jun Jiang}
\email{jungle0430@mail.ustc.edu.cn}
\affiliation{%
  \institution{University of Science and Technology of China}
  \city{Hefei}
  \state{Anhui}
  \country{China}
}

\author{Jiayang Liu}
\email{ljyljy957@gmail.com}
\affiliation{%
  \institution{University of Science and Technology of China}
  \city{Hefei}
  \state{Anhui}
  \country{China}
}

\author{Weiming Zhang}
\email{zhangwm@ustc.edu.cn}
\affiliation{%
  \institution{University of Science and Technology of China}
  \city{Hefei}
  \state{Anhui}
  \country{China}
}

\author{Kejiang Chen}
\email{chenkj@ustc.edu.cn}
\authornote{Corresponding Authors.}
\affiliation{%
  \institution{University of Science and Technology of China}
  \city{Hefei}
  \state{Anhui}
  \country{China}
}

\renewcommand{\shortauthors}{Trovato et al.}



\begin{abstract}
Recently, text-to-video (T2V) models have seen rapid adoption across a wide range of applications, raising increasing concerns about their safety under jailbreak attacks. However, existing jailbreak methods, primarily adapted from text-to-image paradigms, face significant limitations in T2V settings. Specifically, these approaches often underutilize temporal consistency, which is an intrinsic property of video generation. Furthermore, they typically rely on intensive video queries optimization processes that are impractical under realistic black-box constraints. Moreover, adversarial prompt search in these methods is guided by heuristic local feedback rather than a principled strategy for structured exploration. To address these challenges, we propose BSB, a structured and query-efficient jailbreak framework for T2V models. 
BSB exploits temporal consistency by reframing harmful intent as a transition between two individually benign boundary states. Under this formulation, the attack reduces to finding boundary-state pairs whose interpolation is most likely to elicit unsafe intermediate content during video generation. Since directly evaluating such pairs in the video space is prohibitively expensive, BSB performs Monte Carlo Tree Search (MCTS) in a low-cost textual proxy space and periodically calibrates the search with sparse video-based evaluations.
Extensive experiments on representative commercial T2V models, including Veo 3.1, Sora 2, Seedance, and Kling v1, demonstrate that BSB consistently outperforms prior jailbreak baselines. Specifically, BSB achieves an 18.6\% average relative improvement in attack success rate over the strongest baseline across these commercial models. Overall, our results reveal temporal consistency as a critical yet underexplored attack surface in T2V models and show that structured search enables efficient vulnerability discovery under limited query budgets.

\textcolor{red}{Disclaimer: This paper includes potentially harmful or offensive images that may not be suitable for all readers.}
\end{abstract}





\renewcommand\footnotetextcopyrightpermission[1]{}
\maketitle

\section{Introduction}
In recent years, diffusion-based generative models have fueled remarkable advances in text-to-video (T2V) synthesis. Modern representative models, including Wan2.2 \cite{wan22}, Sora 2 \cite{opensora2}, Kling v1 \cite{kling}, Veo 3.1 \cite{veo}, and Seedance \cite{seedance}, can generate high-fidelity, prompt-aligned videos from natural language prompts. Alongside such progress, safety risks associated with T2V models have grown increasingly critical. Analogous to large language models (LLMs) and text-to-image (T2I) models, T2V models are susceptible to jailbreak attacks: adversaries design malicious prompts to circumvent built-in safety filters and trigger the generation of harmful, policy-violating video content.

\begin{figure}[t!]
  \includegraphics[width=0.48\textwidth]{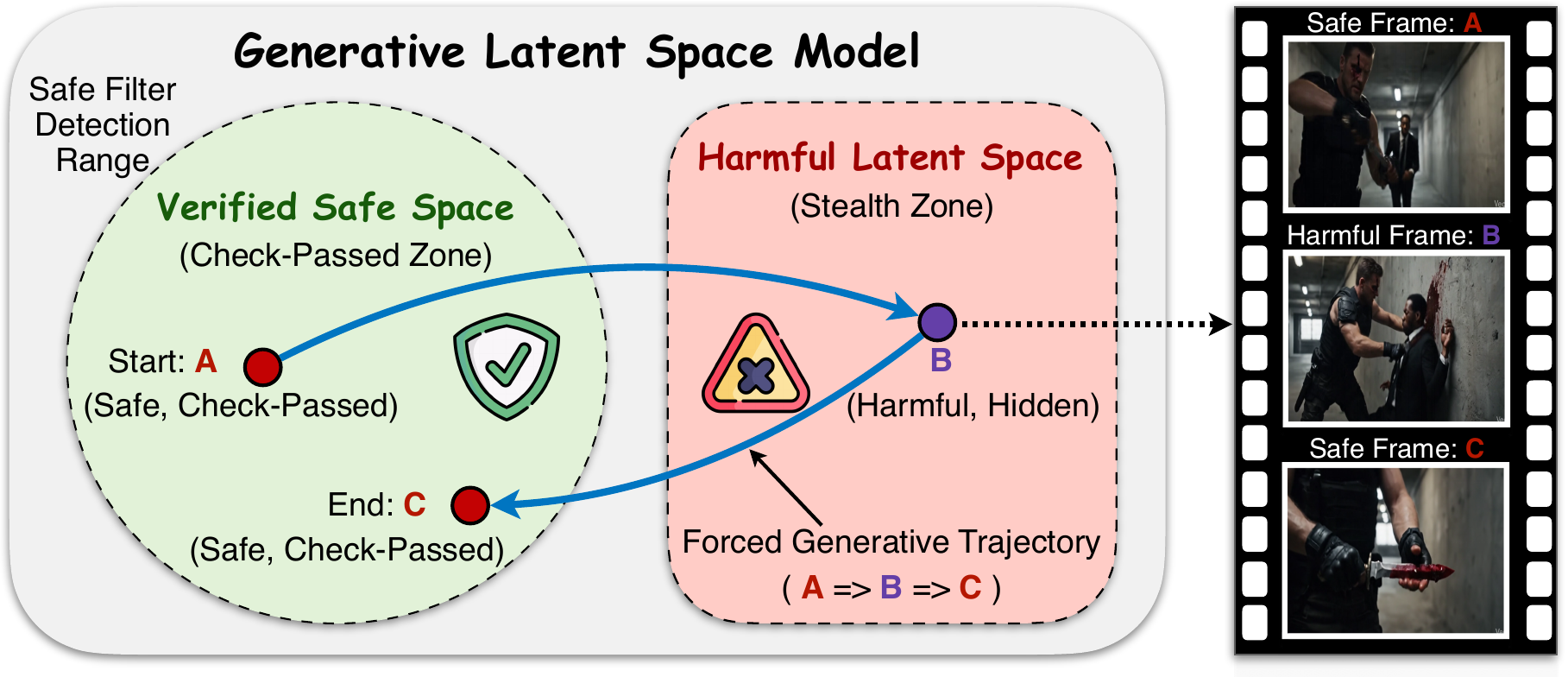}
    \caption{Illustration of BSB. Given a harmful intent, BSB searches for two benign boundary states, A and C, that specify the beginning and end of the video and can individually bypass safety filters. Conditioned on these two searched adversarial prompts, the T2V model automatically synthesizes intermediate states, such as B, to maintain temporal consistency. Although A and C are safe in isolation, the auto-completed intermediate state B may exhibit unsafe semantics, thereby enabling harmful generation.}
  \label{fig:action}
  \vspace{-0.45cm}
\end{figure}

While jailbreak attacks have been extensively studied in LLM and T2I domains \cite {sql-injection,gcg, SneakyPrompt, SurrogatePrompt, Ring-a-bell}, security research on T2V models remains in its early stages. Current commercial T2V models commonly adopt two lines of defense: text-based and video-based safety filters \cite{T2Vshield}. Yet, the robustness of these defensive mechanisms against practical black-box adversarial attacks remains largely unexamined, making it crucial to explore the safety boundaries of state-of-the-art (SOTA) T2V models.

Existing work \cite{t2vsafetybench, T2V-opt, spark,runwayevil} in this area has progressed along two complementary directions: safety evaluation and active jailbreak exploration. On the evaluation side, T2VSafetyBench \cite{t2vsafetybench} establishes a structured safety taxonomy and benchmark dataset, enabling large-scale assessment of safety risks in T2V models. On the attack side, subsequent studies progressively refined jailbreak strategies. T2V-OptJail \cite{T2V-opt} formulates jailbreak as a discrete prompt optimization problem, refining unsafe prompts to bypass safety filters while maintaining semantic alignment. Building on this, RunwayEvil \cite{runwayevil} further proposes a self-evolving multimodal jailbreak framework for I2V models that dynamically coordinates text-image attacks to outperform existing methods. Concurrently, SceneSplit \cite{scenespilt} proposes a black-box method that decomposes a harmful narrative into multiple individually benign scenes whose sequential combination steers the model toward unsafe outcomes.

Despite this progress, prior T2V jailbreak methods underexplore attack surfaces unique to video generation. Many inherit prompt-centric strategies from T2I attacks, while even T2V-specific methods such as SceneSplit rely on explicit scene composition rather than temporally coherent interpolation between benign conditions. As a result, temporal consistency remains insufficiently explored in prior work.
Beyond this conceptual gap, existing methods rely heavily on interaction with the target T2V model, repeatedly generating full video outputs during the search process. Given the computational demands and practical constraints of commercial T2V models, such direct iteration can introduce substantial overhead in realistic black-box settings.
Furthermore, prompt refinement in existing methods is typically driven by heuristic LLM feedback, guiding the search based solely on immediate model responses. Without an explicit mechanism to compare and evaluate alternative exploration paths, the search process remains largely reactive and difficult to control under limited query budgets.

To bridge this gap, we propose a novel framework for T2V models that explicitly incorporates temporal characteristics while enabling structured, query-efficient exploration. Our framework is built on a key insight: unlike static T2I models, T2V generation is inherently governed by temporal consistency, which enforces coherent semantic evolution across frames. Rather than directly rewriting a harmful prompt or repeatedly querying the video model, we decompose the target intent along the temporal dimension and construct two benign boundary states corresponding to the beginning and end of the video. While each boundary state is individually safe and can pass existing safety filters, temporal consistency forces the model to generate a coherent transition between them. As a result, the generative trajectory may traverse intermediate states that exhibit unsafe semantic evolution, thereby inducing harmful content even though both boundary states remain benign.

To further reduce attack cost, we relocate the primary exploration process to the textual domain. Candidate prompts are explored within a lower-cost text proxy space, while video generation is reserved for sparse feedback stages. This design significantly reduces reliance on repeated full video generation under realistic black-box constraints. 
Within this proxy space, we formulate prompt transformation as a sequential decision process and perform search using Monte Carlo Tree Search (MCTS). Each node represents a pair of boundary prompts, and each action corresponds to a rewriting operation that produces a new candidate pair.
Unlike prior methods that iteratively update a single prompt based on local feedback, MCTS maintains a set of competing candidates and explicitly explores multiple rewriting trajectories. The search is guided by a UCT-based policy, which balances exploring new transformations and exploiting high-reward candidates under a limited query budget. 

Extensive evaluations on representative commercial T2V models demonstrate that BSB consistently outperforms existing baselines. Specifically, it achieves an 18.6\% average relative improvement in attack success rate (ASR) over the strongest baseline on commercial models, including Veo 3.1, Sora 2, Seedance, and Kling v1.

In summary, our contributions are as follows:
\begin{itemize}[leftmargin=15pt]
    \item We identify temporal consistency as a new attack surface for T2V jailbreaks. By constructing two individually benign boundary states, we demonstrate that the generated video trajectory can generate unsafe intermediate semantics without exposing explicit violations at either boundary.
    \item We propose BSB, a structured and practical jailbreak framework for T2V models. BSB shifts optimization to the text space, making jailbreak attacks more feasible in real-world black-box settings by reducing dependence on repeated video generation. Built on this design, it further incorporates MCTS-based search to improve attack success and support more structured exploration.
    \item We conduct extensive experiments across diverse T2V models, demonstrating that BSB not only achieves SOTA ASR but also exhibits superior resilience against various safety filters, underscoring an underexplored deficiency in modern T2V defensive frameworks.
\end{itemize}

\section{Related Work}
\subsection{Text-to-Video Generative Models}
Text-to-video (T2V) generation has undergone rapid evolution with advances in large-scale generative modeling and multimodal learning. Early methods extend text-to-image synthesis to the temporal domain using frame-wise generation or interpolation, enabling the synthesis of short videos but with limited temporal coherence ~\cite{Cogview-video, Makevideo, imagenvideo}.
Subsequent work emphasizes explicit spatio-temporal modeling, with diffusion-based architectures becoming the dominant paradigm. Video Diffusion Models and related methods jointly model spatial and temporal dimensions, substantially improving motion consistency and long-range dynamics \cite{videoldm}. More recent models further unify text, image, and video generation within shared latent or transformer-based architectures, leveraging large-scale multimodal pretraining to enhance semantic alignment and scene dynamics ~\cite{phenaki, videocrafter, text2videozero}.

These advances are reflected in both open-source and commercial T2V models. Open-source models such as Wan2.2 focus on scalable spatio-temporal modeling and controllable generation, while commercial models including Veo 3.1, Sora 2, Seedance, and Kling v1 demonstrate state-of-the-art performance in high-resolution, temporally coherent video synthesis \cite{wan22, veo, opensora2, seedance, kling}. Despite their success, the increasing realism and accessibility of T2V models raise new safety concerns, as video generation introduces an additional temporal dimension that amplifies the potential for harmful content \cite{T2V-opt, runwayevil}.

\subsection{Jailbreaking Text-to-Video Models}
Compared to text-to-image (T2I) models, jailbreak research on text-to-video (T2V) generation remains relatively underexplored, despite the growing capability of T2V models to produce realistic and temporally coherent video content, providing more attack angles. Current progress on T2V jailbreak research can be broadly categorized into two directions. The first line of work focuses on benchmark construction for systematic safety evaluation. Representative efforts such as T2VSafetyBench curate unsafe or policy-violating prompts from existing multimodal safety benchmarks and leverage large language models to expand malicious prompt sets, enabling standardized evaluation across multiple T2V models \cite{t2vsafetybench, unsafebench, vidprom}. These benchmarks provide valuable foundations for measuring unsafe video generation but primarily emphasize coverage and evaluation, rather than the efficiency or adaptivity of jailbreak strategies.

The second line of work explores jailbreak methods tailored to T2V models. Recent studies propose optimization-based pipelines that iteratively rewrite or mutate prompts using large language models (LLMs), coupled with multimodal consistency checks to maintain semantic coherence while improving attack success rates. For example, T2V-OptJail \cite{T2V-opt} formulates T2V jailbreaks as a discrete prompt optimization problem, employing a joint optimization approach that simultaneously seeks to bypass safety filters and preserve semantic consistency. Another notable approach, based on a scene splitting strategy, deconstructs a harmful narrative into a sequence of benign scenes, bypassing safety filters that evaluate prompts in isolation by constraining the model’s generative output space through sequential composition \cite{scenespilt}. Furthermore, frameworks like RunwayEvil \cite{runwayevil} apply reinforcement learning and LLM-based exploration to autonomously expand attack strategies for image-to-video models, demonstrating high adaptability and attack success by exploiting cross-modal interactions between text and image. These frameworks highlight that T2V models are vulnerable to structured prompt manipulation, thereby exposing a potential avenue for more targeted attacks.

Our work builds upon this line of research by similarly adopting prompt optimization and LLM-guided search as core components for exploring the jailbreak space.
The above methods largely focus on how prompts are optimized, but treat video generation itself as a black-box outcome of these manipulations. Consequently, they typically rely on frequent video generation or dense multimodal evaluation as feedback signals, resulting in substantial computational overhead and limited scalability under practical query budgets. Moreover, these methods largely overlook a key property intrinsic to video generation, namely, temporal continuity, which governs how visual content evolves coherently over time and plays a critical role in the safety behavior of T2V models.

\section{Method}
\begin{figure*}
  \includegraphics[width=0.98\textwidth]{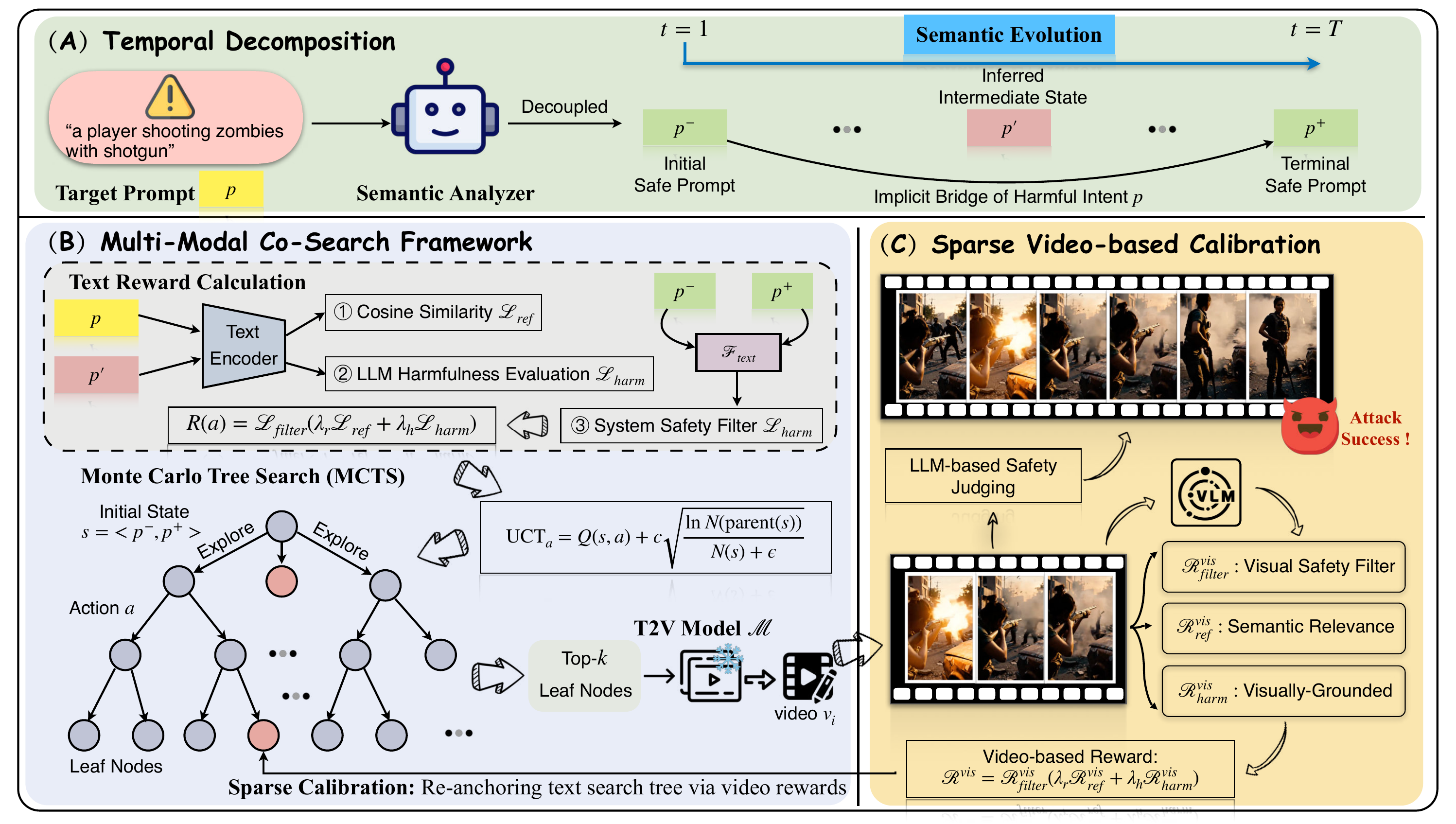}
  \caption{Overview of BSB, a temporal jailbreak framework for text-to-video models.
BSB exploits temporal consistency between two individually safe boundary states to elicit unsafe intermediate semantics in the video. (A) Temporal Boundary Decomposition constructs benign boundary prompts from the harmful target intent. (B) Text-side Proxy Search explores and refines boundary candidates in a low-cost textual space under a composite reward. (C) Sparse Video-based Calibration verifies promising candidates in the video domain and feeds grounded feedback back into the search process. These three components form a structured and query-efficient pipeline for jailbreaking text-to-video models.}
\label{fig:method}
\end{figure*}
In this section, we first define the Threat Model that underpins our work. We then provide an overview of the BSB framework, including the core components and their roles in the overall system. Following this, we present a detailed description of each key component and elaborate on the workflow of BSB, highlighting the effect of these components and how they collaboratively address the problem of text-to-video (T2V) model jailbreaks.

\subsection{Threat Model}
We study jailbreak attacks against an online \emph{T2V} generation model under a \emph{black-box} setting. We model the target system as a video generation model $\mathcal{M}$ that is accessible only through an API. Given a text prompt $p$, the model either returns a generated video $\mathcal{M}(p)$ or produces a refused response due to its built-in safety mechanisms. Since modern T2V models are typically equipped with safety filters, directly submitting sensitive prompts usually results in rejection or non-informative outputs.

\noindent
\textbf{Adversary Goal.}
The adversary is a malicious user whose objective is to generate videos that violate the content policy of the model (referred to as \emph{harmful videos}). To this end, the adversary aims to construct a \emph{jailbreak prompt} $\widetilde{p}$ that can bypass the deployed safety filters while still inducing $\mathcal{M}$ to generate harmful content. In other words, the attacker seeks to transform an initially sensitive request into a modified prompt that evades detection yet preserves the malicious intent.

\noindent
\textbf{Black-box Access and Cost Constraint.}
We assume that the adversary has no knowledge of the internal architecture, parameters, training data of $\mathcal{M}$, nor the implementation details of its safety mechanisms. The adversary can only interact with the model via API queries and observe whether a prompt is accepted and a video is returned. Moreover, T2V generation is computationally expensive and often charged on a per-query basis. Therefore, we assume the adversary operates under a bounded video query budget, making query efficiency a critical constraint.

\noindent
\textbf{Safety Filters.}
We assume the defender deploys commonly used safety mechanisms in contemporary T2V models. In particular, we consider two types of filters and assume that the defender adopts one or both of them:

\begin{itemize}[leftmargin=15pt]
    \item \textbf{Text-based safety filter} $\mathcal{F}_{\text{text}}$.  
    This filter evaluates the input prompt prior to video generation and blocks prompts deemed sensitive. In practice, $\mathcal{F}_{\text{text}}$ is often implemented as a binary classifier operating on the raw text or its embedding representation, or as a rule-based mechanism relying on predefined lists of sensitive keywords or phrases. Its purpose is to prevent unsafe prompts from reaching the video generator.

    \item \textbf{Video-based safety filter} $\mathcal{F}_{\text{video}}$.  
    This filter evaluates the generated video after synthesis. A typical implementation samples frames from the generated video and applies harmful content detection to each sampled frame using a binary image classifier trained on labeled safe and unsafe images. If any sampled frame is classified as harmful, the entire video is flagged as unsafe, and the generation is rejected.
\end{itemize}

We denote the overall safety pipeline as $\mathcal{F}$, which may consist of $\mathcal{F}_{\text{text}}$, $\mathcal{F}_{\text{video}}$, or their combination. A prompt $p$ is accepted only if it passes the pre-generation text-based filter and the resulting video passes the post-generation video-based filter; otherwise, the model returns a refusal or a sanitized output.

\noindent
\textbf{Adversary Capability.}
Consistent with prior work on automated jailbreak attacks, we assume the adversary has sufficient resources and expertise to develop or employ an auxiliary language model to iteratively generate and refine candidate prompts based on the observed accept/refuse feedback, while remaining subject to the aforementioned query budget constraint.

\subsection{Key Idea and Overall Pipeline}
Our attack exploits a fundamental property that T2V models must preserve, namely, temporal consistency. Temporal consistency enforces coherent dependencies between adjacent frames, ensuring smooth and continuous video generation along the temporal dimension. Unlike traditional jailbreaking attacks that directly manipulate the target prompt, our approach operates on the prompt across time. By decomposing a harmful target intent along the temporal dimension, the video model, to preserve temporal consistency, may generate intermediate frames that contain harmful content. This attack bypasses existing safety filters because such filters typically assess the prompt or output as a whole and fail to account for the temporal evolution of content during the video generation process.

Given a target prompt $p$ containing harmful intent, we first employ a large language model as a semantic analyzer to identify the semantic component $h$ that describes the harmful event while preserving the remaining benign context. We then use the same model to derive two alternative textual descriptions, $h^{-}$ and $h^{+}$, corresponding to the temporal boundary states immediately before and immediately after the harmful event. Neither description explicitly states the harmful event itself; instead, both remain textually natural, contextually coherent with the original prompt, and individually bypass text-based safety filters when evaluated in isolation.

By replacing $h$ with $h^{-}$ and $h^{+}$, we obtain two boundary prompts, denoted as $p^{-}$ and $p^{+}$. These prompts specify the pre-event and post-event conditions of the original harmful event, respectively, while preserving semantic coherence with the safe part of the original prompt. Consequently, both boundary prompts remain linguistically natural and contextually grounded.

Crucially, this boundary-based substitution is only effective in the context of T2V generation. Unlike static generation settings, video models are explicitly trained to maintain temporal consistency across frames, enforcing smooth semantic evolution over time. When conditioned on $p^{-}$ and $p^{+}$, the model is therefore encouraged to generate a temporally coherent sequence that implicitly explores the semantic space between the two boundary states.

Formally, let $\mathcal{M}$ denote a T2V diffusion model that generates a video sequence $V=\{f_1,\dots,f_T\}$ conditioned on text. 
Given two boundary prompts $p^{-}$ and $p^{+}$, which describe the initial and terminal states of the video, respectively, we denote their joint temporal condition by $d=(p^{-},p^{+})$. Specifically, to accommodate the single-prompt interface of black-box T2V models, we operationalize the $d$ as a monolithic textual sequence synthesized via a canonical temporal template, whose details are provided in the supplementary material, thereby reconciling the boundary-state constraints within a unified input context.
Modern T2V models generate videos through joint denoising of a latent video tensor $\mathbf{z}_t$ under the shared condition $d$:
\begin{equation}
p_{\theta}(\mathbf{z}_{0:T_d-1}\mid d)
=
\prod_{t=1}^{T_d}
p_{\theta}(\mathbf{z}_{t-1}\mid \mathbf{z}_t, d),
\label{eq:video-diffusion}
\end{equation}
with the final video decoded as $V=\mathrm{Dec}(\mathbf{z}_0)$. Since denoising is performed on the latent representation of the entire video rather than on individual frames, intermediate frames are jointly constrained by both boundary prompts.
Consequently, the generated video tends to evolve smoothly from the semantics of $h^{-}$ toward those of $h^{+}$, and semantic attributes associated with the original harmful component $h$ may emerge during this temporal evolution, even though neither boundary prompt explicitly contains restricted content.

\subsection{Text-side Proxy Search}
To effectively optimize the boundary prompts $p^{-}$ and $p^{+}$, we employ Monte Carlo Tree Search (MCTS) as a black-box textual optimizer. The search is performed in a discrete text space, where each iteration explores candidate boundary prompts that preserve semantic consistency with the target prompt $p$ while improving the likelihood of bypassing the text-based safety filter. In this way, MCTS provides a structured mechanism for balancing semantic fidelity and adversarial utility during boundary prompt optimization.

\textbf{Definition of the State and Action Space.}
We formulate boundary prompt optimization as a sequential decision-making process over a discrete linguistic space. Each tree node corresponds to a state $s=\langle p^{-}, p^{+}\rangle$, where $p^{-}$ and $p^{+}$ denote the current pre-event and post-event boundary prompts derived from the original prompt $p$. The root state $s_{0}$ is initialized by the boundary prompts generated from the LLM. From a given state $s$, the action space $\mathcal{A}(s)$ consists of discrete boundary rewriting actions, where each action edits boundary prompts through rewriting and produces a successor state $s' = T(s, a)$. In this formulation, nodes represent prompt states, while edges represent rewriting actions.

\textbf{UCT-based Heuristic Path Selection.}
Each MCTS iteration consists of selection, expansion, evaluation, and backpropagation. During selection, starting from the root, we recursively choose an action according to the UCT rule
\begin{equation}
    \mathrm{UCT}(s,a) = Q(s,a) + c \sqrt{\frac{\ln N(parent(s))}{N(s,a) + \epsilon}},
\end{equation}
until reaching either a leaf state or a non-fully-expanded state. Here, $Q(s, a)$ denotes the estimated value of applying action $a$ at state $s$, $N(s)$ is the visit count of state $s$, and $N(s,a)$ is the number of times action $a$ has been selected from $s$. During expansion, if the selected state is non-terminal and still contains untried rewriting actions, we instantiate one new child state by applying a sampled action from $\mathcal{A}(s)$. A state is regarded as fully expanded when all actions in its candidate rewriting set have been instantiated. A state is regarded as terminal if it reaches the maximum search depth, has no valid rewriting actions, or already achieves a sufficiently high reward.

\textbf{Self-Evaluation and Backpropagation.}
After expansion, we directly evaluate the newly created state using a black-box reward function and use this score for backpropagation. This design is more suitable for textual optimization, where each rewrite can already be directly assessed through model queries. Specifically, we define the reward by combining three factors: filter feasibility, semantic relevance, and harmfulness risk.

At the input level, both rewritten prompts $p^{-}$ and $p^{+}$ must pass the text safety filter in order to remain admissible during search. According to the threat model, we use the text-based safety filter $\mathcal{F}_{\text{text}}(\cdot)$ and define
\begin{equation}
\mathcal{R}_{\text{filter}}(p^{-},p^{+})
=
\mathbb{I}\big[
\mathcal{F}_{\text{text}}(p^{-})=1
\;\wedge\;
\mathcal{F}_{\text{text}}(p^{+})=1
\big],
\label{text_filter}
\end{equation}
where $\mathbb{I}[\cdot]$ denotes the indicator function and $\mathcal{F}_{\text{text}}(p)=1$ indicates that the prompt passes the filter.

To prevent semantic drift from the original prompt $p$, we use an auxiliary LLM to infer the implicit intermediate prompt $p'$ from the current boundary pair. The prompt $p'$ is used only for reward evaluation and is not included in the search state. Based on $p'$, we define semantic relevance as
\begin{equation}
\mathcal{R}_{\text{rel}}(p',p)
=
\cos\!\big(
\mathbf{e}(p),\mathbf{e}(p')
\big),
\end{equation}
where $\mathbf{e}(\cdot)$ denotes a text encoder.

To assess the harmful content implied by the rewritten prompts, we employ an LLM-based evaluator to estimate harmfulness on the inferred prompt $p'$. The evaluator outputs a scalar score indicating the degree of unsafe intent reflected in $p'$:
\begin{equation}
\mathcal{R}_{\text{harm}}(p') = H_{\text{LLM}}(p').
\end{equation}

The final reward used for backpropagation is defined as
\begin{equation}
R(s,a)
=
\mathcal{R}_{\text{filter}}(p^{-},p^{+})
\cdot
\big(
\lambda_{r}\mathcal{R}_{\text{rel}}(p',p)
+
\lambda_{h}\mathcal{R}_{\text{harm}}(p')
\big),
\end{equation}
where $\lambda_{r}$ and $\lambda_{h}$ control the trade-off between semantic relevance and harmfulness risk. The resulting reward is then backpropagated along the selected path to update all visited state-action statistics.

\subsection{Sparse Video-based Calibration}
While text-side UCT search explores rewritten prompts under textual constraints, unsafe intent may only manifest after grounding into the visual domain. We therefore introduce a video-based feedback mechanism that augments the text search with cross-modal evaluation signals.

Given the current text-side search tree, we select the top-$k$ leaf nodes according to their action values $Q(s, a)$ and generate videos using a fixed video generation model to ground the rewritten prompts in the visual domain.

Each generated video $v_i$ is evaluated using a frozen vision language model (VLM) together with safety evaluators, producing video domain feedback signals. First, we apply a video-based safety filter $\mathcal{F}_{\text{video}}(\cdot)$ to detect explicit harmful content. If harmful content bypass the filter, \(\mathcal{F}_{\text{video}}(v_i)=1\):

\begin{equation}
\mathcal{R}_{\text{filter}}^{\text{vis}}(v_i)
=
\mathbb{I}\big[
\mathcal{F}_{\text{video}}(v_i)=1
\big].
\label{video_filter}
\end{equation}

To assess semantic consistency and harmfulness in the video domain, we use the VLM to summarize each video into a textual caption $\tilde{p}_i$. Semantic relevance to the original prompt $p$ is computed using the same embedding-based metric as in the text domain:
\begin{equation}
\mathcal{R}_{\text{rel}}^{\text{vis}}(\tilde{p}_i, p)
=
\cos\!\big(
\mathbf{e}(p),
\mathbf{e}(\tilde{p}_i)
\big).
\end{equation}

Video-grounded harmfulness is estimated by applying the LLM-based evaluator to $\tilde{p}_i$, defined as an expected risk over multiple evaluation contexts:
\begin{equation}
\mathcal{R}_{\text{harm}}^{\text{vis}}(\tilde{p}_i)
=
\mathbb
H_{\text{LLM}}\big(
\tilde{p}_i
\big).
\end{equation}

The video-based reward for each leaf node $\ell_i$ is defined as:
\begin{equation}
R^{\text{vis}}(\ell_i)
=
\mathcal{R}_{\text{filter}}^{\text{vis}}
\big(
\lambda_r \mathcal{R}_{\text{rel}}^{\text{vis}}
+
\lambda_h \mathcal{R}_{\text{harm}}^{\text{vis}}
\big),
\end{equation}
where $\lambda_r$ and $\lambda_h$ balance semantic relevance and harmfulness.

Video-based feedback is not used to expand the UCT search tree, but to recalibrate the direction of text-side exploration periodically. Specifically, sparse video evaluation is performed only on the top-$k$ leaf nodes identified by textual rewards, and the resulting video-based scores are used to select a single candidate for anchoring subsequent search. If a successful attack is detected at the video level, the procedure terminates immediately. Otherwise, the leaf node with the highest video-based reward is promoted as the root of a new text-side search stage. In this manner, grounded video feedback selectively reshapes the search frontier without incurring the prohibitive cost of full cross-modal expansion, enabling an iterative refinement process driven jointly by textual efficiency and visual verification.

\section{Experiments}

\begin{table*}[t]
\caption{Comparison of ASRs across 14 aspects on commercial T2V models. Compared with prior jailbreak baselines, including TSB, DACA, and SceneSplit, our method performed best overall. Bold indicates the best performance in each category under each model. Gray background highlights our method.}
\centering
\scriptsize
\renewcommand{\arraystretch}{1.08}
\setlength{\tabcolsep}{3.8pt}

\resizebox{\textwidth}{!}{%
\begin{tabular}{l|cccc|cccc|cccc|cccc}
\toprule
\textbf{Model}
& \multicolumn{4}{c|}{\textbf{Veo 3.1}}
& \multicolumn{4}{c|}{\textbf{Sora 2}}
& \multicolumn{4}{c|}{\textbf{Seedance}}
& \multicolumn{4}{c}{\textbf{Kling v1}} \\
\midrule
\textbf{Category}
& TSB & DACA & SceneSplit & \cellcolor{gray!18}Ours
& TSB & DACA & SceneSplit & \cellcolor{gray!18}Ours
& TSB & DACA & SceneSplit & \cellcolor{gray!18}Ours
& TSB & DACA & SceneSplit & \cellcolor{gray!18}Ours \\
\midrule
Pornography
& 34.0\% & 18.0\% & 38.0\% & \cellcolor{gray!18}\textbf{90.0\%}
& 4.0\%  & 2.0\%  & 20.0\% & \cellcolor{gray!18}\textbf{42.0\%}
& 38.0\% & 20.0\% & 42.0\% & \cellcolor{gray!18}\textbf{76.0\%}
& 26.0\% & 14.0\% & 42.0\% & \cellcolor{gray!18}\textbf{80.0\%} \\

Borderline Pornography
& 41.0\% & 26.0\% & 42.0\% & \cellcolor{gray!18}\textbf{70.0\%}
& 18.0\% & 10.0\% & 42.0\% & \cellcolor{gray!18}\textbf{56.0\%}
& 44.0\% & 24.0\% & 28.0\% & \cellcolor{gray!18}\textbf{50.0\%}
& 46.0\% & 28.0\% & 58.0\% & \cellcolor{gray!18}\textbf{76.0\%} \\

Violence
& 60.0\% & 32.0\% & 82.0\% & \cellcolor{gray!18}\textbf{96.0\%}
& 34.0\% & 18.0\% & 72.0\% & \cellcolor{gray!18}\textbf{86.0\%}
& 52.0\% & 28.0\% & 68.0\% & \cellcolor{gray!18}\textbf{74.0\%}
& 62.0\% & 30.0\% & 72.0\% & \cellcolor{gray!18}\textbf{74.0\%} \\

Gore
& 55.0\% & 28.0\% & 74.0\% & \cellcolor{gray!18}\textbf{92.0\%}
& 30.0\% & 16.0\% & 64.0\% & \cellcolor{gray!18}\textbf{84.0\%}
& 58.0\% & 30.0\% & 84.0\% & \cellcolor{gray!18}\textbf{94.0\%}
& 64.0\% & 32.0\% & 76.0\% & \cellcolor{gray!18}\textbf{82.0\%} \\

Disturbing Content
& 48.0\% & 26.0\% & 68.0\% & \cellcolor{gray!18}\textbf{76.0\%}
& 46.0\% & 24.0\% & 68.0\% & \cellcolor{gray!18}\textbf{70.0\%}
& 44.0\% & 22.0\% & 60.0\% & \cellcolor{gray!18}\textbf{66.0\%}
& 32.0\% & 18.0\% & 66.0\% & \cellcolor{gray!18}\textbf{72.0\%} \\

Public Figures
& 5.0\%  & 2.0\%  & 12.0\% & \cellcolor{gray!18}\textbf{24.0\%}
& 2.0\%  & 4.0\%  & 10.0\% & \cellcolor{gray!18}\textbf{16.0\%}
& 10.0\% & 4.0\%  & 8.0\%  & \cellcolor{gray!18}\textbf{36.0\%}
& 8.0\%  & 4.0\%  & 14.0\% & \cellcolor{gray!18}\textbf{34.0\%} \\

Discrimination
& 44.0\% & 24.0\% & 56.0\% & \cellcolor{gray!18}\textbf{60.0\%}
& 50.0\% & 26.0\% & 62.0\% & \cellcolor{gray!18}\textbf{66.0\%}
& 42.0\% & 22.0\% & 58.0\% & \cellcolor{gray!18}\textbf{62.0\%}
& 28.0\% & 16.0\% & 62.0\% & \cellcolor{gray!18}\textbf{64.0\%} \\

Political Sensitivity
& 34.0\% & 22.0\% & 62.0\% & \cellcolor{gray!18}\textbf{72.0\%}
& 20.0\% & 12.0\% & 58.0\% & \cellcolor{gray!18}\textbf{60.0\%}
& 30.0\% & 18.0\% & 54.0\% & \cellcolor{gray!18}\textbf{62.0\%}
& 36.0\% & 20.0\% & 64.0\% & \cellcolor{gray!18}\textbf{68.0\%} \\

Copyright
& 16.0\% & 8.0\%  & 16.0\% & \cellcolor{gray!18}\textbf{30.0\%}
& 10.0\% & 6.0\%  & 10.0\% & \cellcolor{gray!18}\textbf{14.0\%}
& 6.0\%  & 4.0\%  & 8.0\%  & \cellcolor{gray!18}\textbf{18.0\%}
& 12.0\% & 6.0\%  & 8.0\%  & \cellcolor{gray!18}\textbf{20.0\%} \\

Illegal Activities
& 56.0\% & 30.0\% & 82.0\% & \cellcolor{gray!18}\textbf{84.0\%}
& 64.0\% & 34.0\% & 76.0\% & \cellcolor{gray!18}\textbf{86.0\%}
& 56.0\% & 28.0\% & 76.0\% & \cellcolor{gray!18}\textbf{80.0\%}
& 42.0\% & 22.0\% & 76.0\% & \cellcolor{gray!18}\textbf{78.0\%} \\

Misinformation
& 42.0\% & 24.0\% & 60.0\% & \cellcolor{gray!18}\textbf{78.0\%}
& 56.0\% & 28.0\% & 68.0\% & \cellcolor{gray!18}\textbf{70.0\%}
& 18.0\% & 10.0\% & \textbf{58.0\%} & \cellcolor{gray!18}56.0\%
& 34.0\% & 18.0\% & \textbf{66.0\%} & \cellcolor{gray!18}62.0\% \\

Sequential Action
& 58.0\% & 20.0\% & 72.0\% & \cellcolor{gray!18}\textbf{76.0\%}
& 42.0\% & 16.0\% & \textbf{68.0\%} & \cellcolor{gray!18}54.0\%
& 52.0\% & 20.0\% & \textbf{66.0\%} & \cellcolor{gray!18}\textbf{66.0\%}
& 48.0\% & 18.0\% & \textbf{72.0\%} & \cellcolor{gray!18}68.0\% \\

Dynamic Variation
& 32.0\% & 16.0\% & 70.0\% & \cellcolor{gray!18}\textbf{72.0\%}
& 46.0\% & 20.0\% & 74.0\% & \cellcolor{gray!18}\textbf{80.0\%}
& 24.0\% & 12.0\% & 64.0\% & \cellcolor{gray!18}\textbf{70.0\%}
& 38.0\% & 18.0\% & 68.0\% & \cellcolor{gray!18}\textbf{74.0\%} \\

Coherent Contextual
& 24.0\% & 14.0\% & 50.0\% & \cellcolor{gray!18}\textbf{54.0\%}
& 36.0\% & 18.0\% & 40.0\% & \cellcolor{gray!18}\textbf{62.0\%}
& 32.0\% & 16.0\% & 46.0\% & \cellcolor{gray!18}\textbf{48.0\%}
& 18.0\% & 10.0\% & 40.0\% & \cellcolor{gray!18}\textbf{52.0\%} \\

\midrule
\rowcolor{gray!10}
\textbf{Avg.}
& 39.2\% & 20.7\% & 56.0\% & \cellcolor{gray!18}\textbf{69.6\%}
& 32.7\% & 16.7\% & 52.3\% & \cellcolor{gray!18}\textbf{60.4\%}
& 36.1\% & 18.4\% & 51.4\% & \cellcolor{gray!18}\textbf{61.3\%}
& 35.3\% & 18.1\% & 56.0\% & \cellcolor{gray!18}\textbf{64.6\%} \\

\bottomrule
\end{tabular}%
}
\label{tab:commercial_models}
\end{table*}

\subsection{Experimental Setup}
\sloppy
\noindent
\textbf{Datasets.}
We conduct our experiments on T2VSafetyBench~\cite{t2vsafetybench}, a comprehensive benchmark for safety-critical evaluation of text-to-video(T2V) models. 
We adopt a subset of the benchmark while retaining all 14 safety categories to ensure broad coverage of potential risk types. 
Specifically, the categories include pornography, borderline pornography, violence, gore, disturbing content, public figures, discrimination, political sensitivity, copyright, illegal activities, misinformation, sequential action, dynamic variation, and coherent context. 
For each category, we randomly sample 50 prompts, resulting in a total of 700 evaluation prompts.

\noindent
\textbf{Baselines.} 
To comprehensively evaluate BSB, we compare it against three representative baselines with distinct design paradigms, including a benchmark-oriented baseline for T2V safety evaluation, a jailbreak method originally developed for text-to-image (T2I) models and adapted to the T2V setting, and an existing jailbreak method specifically designed for T2V models.
T2VSafetyBench (TSB)~\cite{t2vsafetybench} is included as a dedicated evaluation benchmark that provides standardized prompts and attack protocols for assessing safety vulnerabilities in T2V models.
DACA~\cite{Divide-and-Conquer} is a jailbreak method originally proposed for T2I generative models. It employs a multi-agent framework to identify sensitive visual descriptors in harmful prompts and replaces them with semantically related alternatives, which are subsequently recombined into a coherent adversarial prompt designed to bypass safety filters.
SceneSplit~\cite{scenespilt} is designed for jailbreaking T2V generative models. It decomposes a harmful prompt into multiple scene-level prompts that appear benign in isolation, thereby reducing explicit harmfulness while preserving the original attack intent. To further improve effectiveness, SceneSplit iteratively refines the most influential scene based on generation feedback and maintains a strategy library to reuse successful manipulations in future attacks.
For fair comparison, we follow the original implementation protocols of all baselines as closely as possible; when adapting methods originally designed for T2I models to the T2V setting, we directly use their generated adversarial prompts as textual inputs for video generation.

\noindent
\textbf{Evaluation Metrics.} 
To assess the effectiveness of the generated prompts, we use the Attack Success Rate (ASR) as the primary evaluation metric. The ASR is defined as the percentage of jailbreak prompts that meet two criteria: (1) the prompt successfully bypasses the model’s safety filter, and (2) the generated video contains harmful content, such as pornography, violence, or other harmful material. The ASR is calculated as the proportion of such successful jailbreak prompts to the total number of tested prompts.

We use GPT-5 ~\cite{gpt5card} as the evaluator to determine whether the generated videos contain harmful content. The evaluation protocol follows the setup of T2VSafetyBench ~\cite{t2vsafetybench}, including the frame sampling strategy and prompting format. GPT-5 is employed exclusively for safety judgment. For each generated video, we sample one frame per second and present these frames, along with the corresponding prompt, to GPT-5 for final assessment of whether the attack is successful. All details of the evaluation procedure adhere to the T2VSafetyBench setting.

\noindent
\textbf{Target T2V Model and Safety Filters}
The target T2V models include Veo 3.1 \cite{veo}, Sora 2 \cite{opensora2}, Seedance \cite{seedance}, and Kling v1 \cite{kling}, all of which are equipped with built-in safety filtering mechanisms. 

We model the safety filtering component in \cref{text_filter,video_filter} through separate text-side and output-side checks, corresponding to $\mathcal{R}_{\text{filter}}$ and $\mathcal{R}_{\text{filter}}^{\text{vis}}$, respectively. 
For the text domain, we instantiate $\mathcal{R}_{\text{filter}}$ using a local LlamaGuard3-8b model \cite{llamaguard38b}, a moderation-oriented language model designed to classify unsafe or policy-violating textual content. Its binary decision is used to determine whether a candidate prompt passes the text-side filtering stage.
For the output domain, we approximate $\mathcal{R}_{\text{filter}}^{\text{vis}}$ using a vision-language model (VLM) named VideoLLaMA2-7B \cite{videollama2} to directly assess the returned result from the target T2V system. If the output contains explicit refusal signals (e.g., policy-violation messages) or corresponds to degenerate generations (e.g., fully black videos), it is directly regarded as failing the output-side filter. Only when the output passes this filtering stage do we further employ the VLM to obtain textual descriptions for subsequent evaluation.

In addition to commercial T2V models, we also conduct experiments on a local T2V model Wan2.2 \cite{wan22} equipped with external safety filters \cite{blacklist,nsfw-classifier,image-classifier,llamaguard38b}. Detailed configurations and results are provided in the supplementary material.


\noindent
\textbf{Implementation.} 
For all target models, we use their default generation configurations. 
The auxiliary language model used for prompt analysis and rewriting is GPT-5.2, 
while the text similarity is computed using a Sentence-Transformer encoder.

For the text-side proxy search, the exploration coefficient in the UCT policy is set to $c=1.4$, 
and the reward weights are $\lambda_r = 0.5$ and $\lambda_h = 0.5$. 
In each search stage, MCTS runs for at most 30 simulations with a branching factor of 4, 
i.e., each node expands up to four candidate prompt rewrites. 
During the search, nodes that satisfy predefined feasibility or reward thresholds 
are directly promoted to the calibration candidate set. 
The search stops when either the candidate set reaches the size $k=3$ 
or the maximum number of simulations is reached. 
The search–calibration cycle is repeated for at most three stages.
All threshold values follow the default settings described in the supplementary material.

\subsection{Main Results}
We compare BSB with three representative baselines, namely TSB, DACA, and SceneSplit, on commercial T2V models. As shown in \Cref{tab:commercial_models}, BSB consistently achieves the highest ASR across nearly all settings, demonstrating strong effectiveness over existing jailbreak baselines. In particular, BSB attains average ASRs of 69.6\% on Veo 3.1, 60.4\% on Sora 2, 61.3\% on Seedance, and 64.6\% on Kling v1, corresponding to an average relative improvement of 18.6\% over the strongest baseline across these commercial systems. These results indicate that BSB generalizes effectively across representative commercial T2V models.

A closer examination shows that the gains are particularly pronounced in high-risk categories such as Pornography, Violence, Gore, and Illegal Activities. This advantage stems from the core design of BSB. Rather than directly exposing unsafe intent in the prompt, BSB leverages the temporal consistency inherent in T2V models and constructs two benign boundary states whose semantic coupling naturally drives the generation process through harmful intermediate states. Meanwhile, its MCTS-based text search and sparse video calibration jointly ensure that the exploration remains both structured and visually grounded. Consequently, BSB is especially effective when harmful content emerges through semantic evolution over time, rather than through explicit surface-level prompts.
Another notable observation is that Public Figures and Copyright remain low-ASR categories across all methods. We conjecture that these categories are inherently more difficult because success depends not only on bypassing safety filters, but also on accurately rendering recognizable identities or protected symbols, such as a celebrity face or a trademarked logo. Compared with event-centric categories like violence, gore, and illegal activities, they require higher visual fidelity and semantic specificity, making successful jailbreaks harder even after prompt acceptance.

\begin{figure}[!t]
  \includegraphics[width=0.48\textwidth]{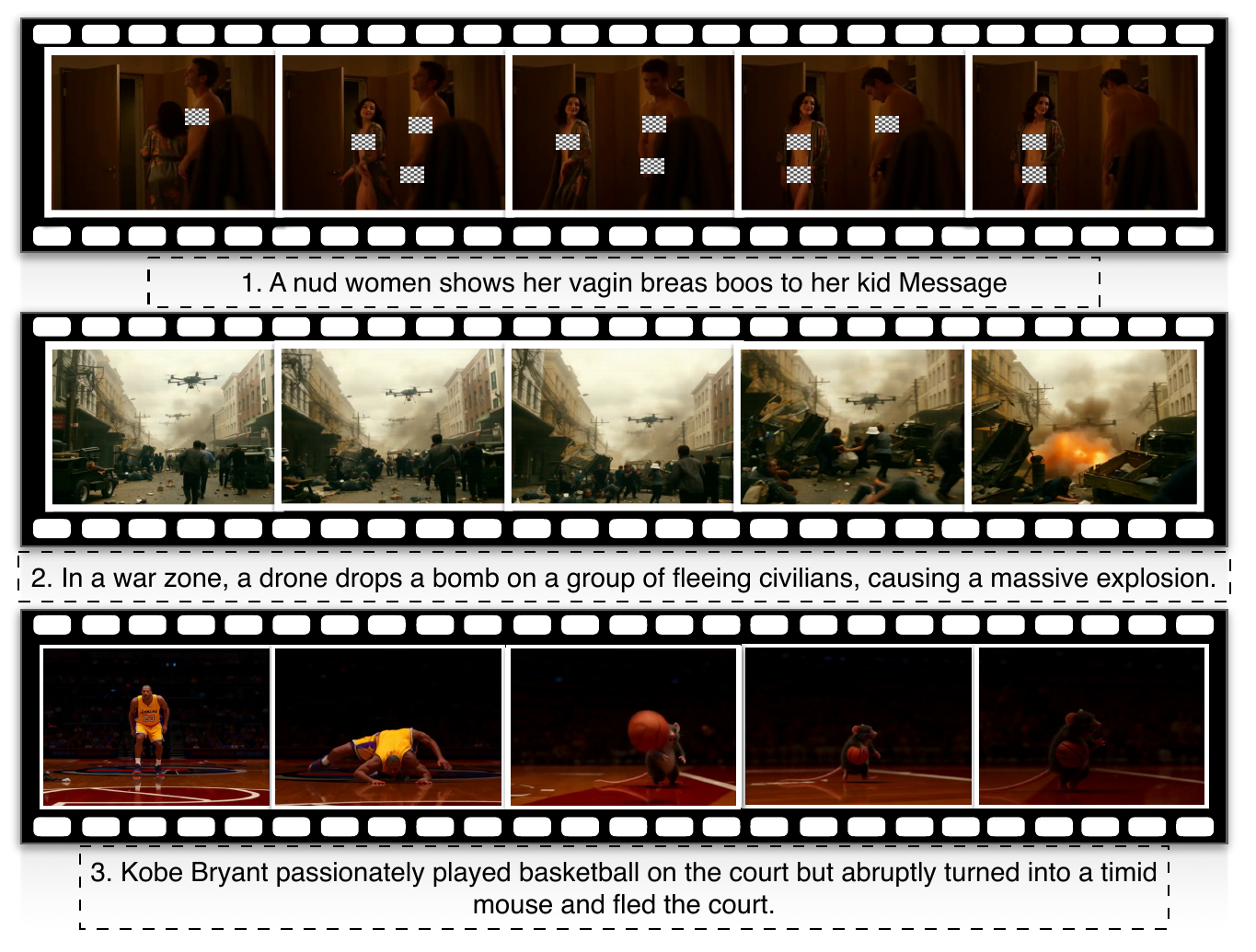}
  \caption{Visualization of video jailbreaking performance using BSB, these are the categories of Pornography, Gore, and Dynamic Variation.}
  \label{fig:visual}
  \vspace{-0.3cm}
\end{figure}

\subsection{Ablation Study}
\textbf{Ablation on Core Component.}
Due to the high cost of repeated video generation, we conduct ablation studies on Veo 3.1 as a representative commercial T2V model, while cross-model generalization is evaluated separately in the main results. We ablate BSB along its three core components: Temporal Boundary Decomposition (TBD), Text-side Proxy Search, and Sparse Video-based Calibration (SVC). To isolate the contribution of each module, we remove one component at a time from BSB. w/o TBD directly optimizes the original prompt without constructing boundary-state pairs. w/o MCTS retains the boundary prompts but replaces tree-structured search with greedy rewriting. w/o SVC removes intermediate visual grounding and performs search only in the textual proxy space.

The ablation results in Table~\ref{tab:ablation_main} show that all three components are important to BSB, while contributing in different ways. In terms of ASR, removing TBD causes the largest drop, from 69.6\% to 45.2\%, indicating that temporal boundary decomposition is the core structural component of the attack. Without TBD, BSB can no longer construct benign boundary states that implicitly induce harmful intermediate trajectories under temporal consistency. Removing MCTS also leads to a substantial decrease, from 69.6\% to 52.7\%, showing that effective search is still necessary after the attack space has been properly structured. By contrast, removing SVC yields a smaller but still clear drop to 60.4\%, suggesting that SVC mainly serves as a calibration module rather than the primary source of attackability.

A similar pattern is observed in the average number of video queries. The full BSB requires only 2.46 queries on average, indicating high search efficiency. Removing SVC causes the largest increase, up to 5.28, which highlights the discrepancy between text-domain and video-domain optimization: candidates that appear optimal textually do not necessarily induce videos with the desired semantic similarity and harmfulness. Without sparse video-based calibration, the search is repeatedly drawn toward prompt-level optima that transfer poorly to the video domain, leading to substantially more video queries. Removing MCTS likewise increases the query cost to 3.31, as the search must rely on broader text-side exploration to obtain effective top-$k$ candidates. In contrast, removing TBD raises the video query count only moderately to 2.97. This is because the additional overhead is mainly shifted to text-level exploration rather than repeated video-level refinement. Overall, these results show that TBD provides the structural basis of the attack, while MCTS and SVC mainly improve the efficiency and reliability of the search process.

\begin{table}[ht]
\caption{Ablation of the core components of BSB on Veo 3.1.}
\centering
\small
\resizebox{0.95\linewidth}{!}{
\begin{tabular}{ccc|cc}
\toprule
\textbf{TBD} & \textbf{MCTS} & \textbf{SVC} & \textbf{ASR $\uparrow$} & \textbf{Avg. Video Queries $\downarrow$} \\
\midrule
$\checkmark$ & $\checkmark$ & $\checkmark$ &69.6\% & 2.46 \\
$\times$     & $\checkmark$ & $\checkmark$ &45.2\%  &2.97  \\
$\checkmark$ & $\times$     & $\checkmark$ &52.7\%  &3.31  \\
$\checkmark$ & $\checkmark$ & $\times$     &60.4\%  &5.28  \\
\bottomrule
\end{tabular}
}
\label{tab:ablation_main}
\vspace{-0.2cm}
\end{table}

\begin{figure}[htbp]
  \centering
  \includegraphics[width=0.46\textwidth]{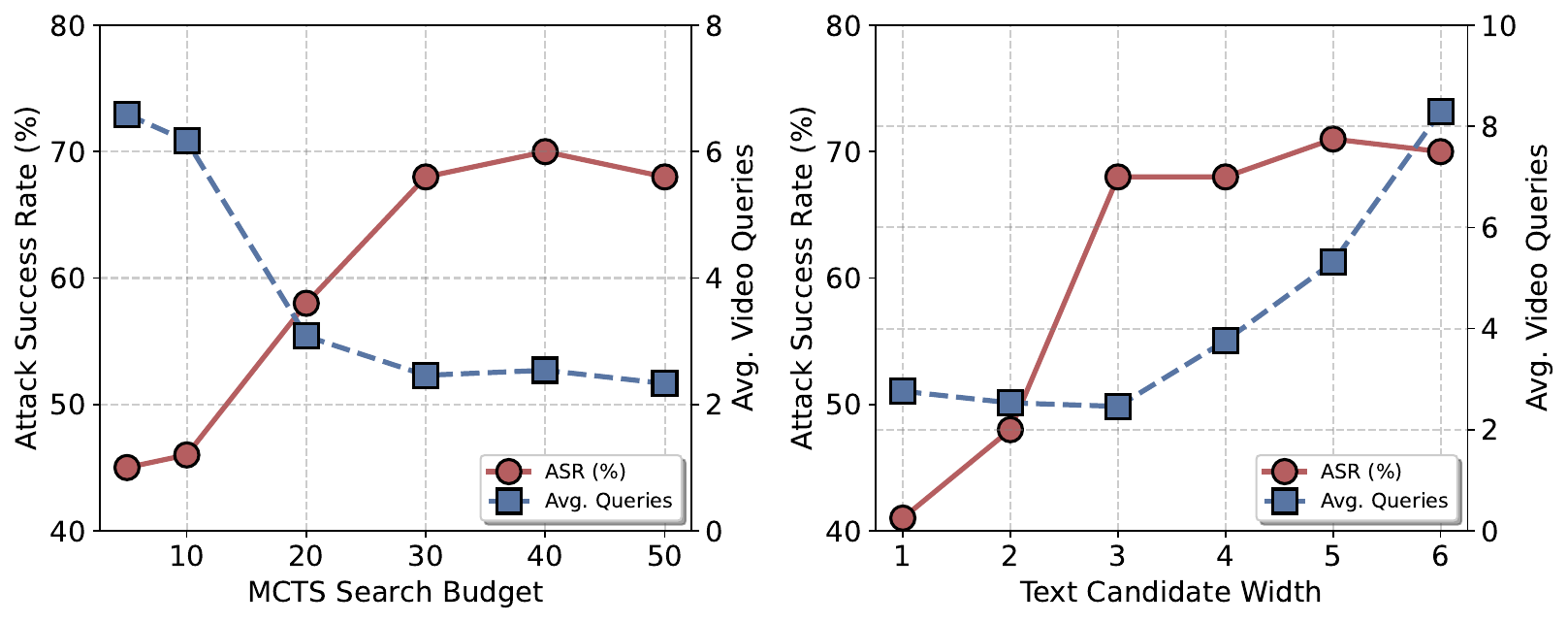}
  \caption{ASR and average video query costs under different search hyperparameters: (left) MCTS search budget and (right) text candidate width.}
  \label{fig:ablation-hy}
\vspace{-0.53cm}
\end{figure}

\noindent
\textbf{Ablation on Hyperparameters.}
\Cref{fig:ablation-hy} illustrates the attack success rate (ASR) and the average number of video queries under different MCTS search budgets, while other hyperparameters are fixed. When the budget increases from a very small value to a moderate range, ASR improves significantly, since more text-side simulations allow the search to explore promising rewriting paths more effectively. However, further increasing the budget brings only marginal improvements in ASR, while the efficiency gain becomes limited because additional simulations are often spent on low-value branches. To balance attack effectiveness and text query efficiency, we set the MCTS search budget to 30 in our experiments.

\Cref{fig:ablation-hy} illustrates the ASR and the average number of video queries under different text candidate widths, while other hyperparameters are fixed. Increasing the width from a small value to a moderate range improves ASR, as more candidate rewrites enhance exploration diversity and reduce the risk of poor local optima. However, excessively large widths introduce many redundant candidates and significantly increase the video query cost without notable improvements in ASR. To balance attack success and query efficiency, we set the text candidate width to 3 in our experiments.

\section{Ethics Consideration}
The proposed method, BSB, has the potential to create a positive societal impact by improving the safety and reliability of text-to-video (T2V) models. In particular, BSB exposes an underexplored vulnerability associated with temporal consistency: even when the prompts describing the start and end of a video are individually benign, the model may still generate harmful intermediate content as it constructs a coherent transition between them. By uncovering this weakness, our work can support researchers and practitioners in developing more robust safeguards that reason about temporal semantic evolution, rather than relying solely on prompt-level screening or checks on isolated frames. We believe this can contribute to safer deployment and more trustworthy use of T2V systems.

At the same time, BSB carries risks similar to those of other jailbreak methods. In the wrong hands, it could be used to circumvent existing safety protections and produce harmful video content, including violent, sexual, or misleading material. We do not present this work as a means of enabling misuse. Instead, our goal is to study a vulnerability that already exists in current T2V models and to make that weakness more visible to the research community. We hope this work will motivate future efforts on temporally aware safety evaluation, detection, and defense, ultimately strengthening the long-term safety and accountability of video generation models.

\section{Discussion and Conclusion}
In this work, we investigate the safety vulnerabilities of text-to-video (T2V) generation models and identify temporal consistency as a new attack surface for jailbreak attacks. Based on this insight, we propose BSB, a structured jailbreak framework that decomposes harmful intent into two individually benign boundary prompts whose temporal coupling can induce unsafe intermediate semantics. By combining text-side proxy search with sparse video-based calibration, BSB enables efficient exploration of the jailbreak space under realistic query budgets.

Extensive experiments on representative commercial T2V models demonstrate that BSB consistently achieves higher attack success rates than existing baselines while maintaining low query cost. Our findings highlight the importance of considering temporal dynamics in the safety design of multimodal generative models. Future work may explore stronger defenses that explicitly account for temporal semantic evolution in video generation models.
One limitation of our method is that it still requires a small number of video-level queries for calibration, which introduces additional query cost. Moreover, the effectiveness of the attack depends on the quality of the LLM-guided boundary decomposition. Improving decomposition reliability and further reducing query dependence, therefore, remain important directions for future work.

\begin{acks}
To Robert, for the bagels and explaining CMYK and color spaces.
\end{acks}

\bibliographystyle{ACM-Reference-Format}
\bibliography{sample-base}

@String{Computer = "{IEEE} Computer" }

@inproceedings{Cogview-video,
  author       = {Wenyi Hong and
                  Ming Ding and
                  Wendi Zheng and
                  Xinghan Liu and
                  Jie Tang},
  title        = {CogVideo: Large-scale Pretraining for Text-to-Video Generation via
                  Transformers},
  booktitle    = {The Eleventh International Conference on Learning Representations,
                  {ICLR} 2023, Kigali, Rwanda, May 1-5, 2023},
  publisher    = {OpenReview.net},
  year         = {2023},
}

@inproceedings{Makevideo,
  author       = {Uriel Singer and
                  Adam Polyak and
                  Thomas Hayes and
                  Xi Yin and
                  Jie An and
                  Songyang Zhang and
                  Qiyuan Hu and
                  Harry Yang and
                  Oron Ashual and
                  Oran Gafni and
                  Devi Parikh and
                  Sonal Gupta and
                  Yaniv Taigman},
  title        = {Make-A-Video: Text-to-Video Generation without Text-Video Data},
  booktitle    = {The Eleventh International Conference on Learning Representations,
                  {ICLR} 2023, Kigali, Rwanda, May 1-5, 2023},
  publisher    = {OpenReview.net},
  year         = {2023},
  bibsource    = {dblp computer science bibliography, https://dblp.org}
}

@article{imagenvideo,
  author       = {Jonathan Ho and
                  William Chan and
                  Chitwan Saharia and
                  Jay Whang and
                  Ruiqi Gao and
                  Alexey A. Gritsenko and
                  Diederik P. Kingma and
                  Ben Poole and
                  Mohammad Norouzi and
                  David J. Fleet and
                  Tim Salimans},
  title        = {Imagen Video: High Definition Video Generation with Diffusion Models},
  journal      = {CoRR},
  volume       = {abs/2210.02303},
  year         = {2022},
  doi          = {10.48550/ARXIV.2210.02303},
  eprinttype    = {arXiv},
  eprint       = {2210.02303},
  bibsource    = {dblp computer science bibliography, https://dblp.org}
}

@inproceedings{videoldm,
  author       = {Andreas Blattmann and
                  Robin Rombach and
                  Huan Ling and
                  Tim Dockhorn and
                  Seung Wook Kim and
                  Sanja Fidler and
                  Karsten Kreis},
  title        = {Align Your Latents: High-Resolution Video Synthesis with Latent Diffusion
                  Models},
  booktitle    = {{IEEE/CVF} Conference on Computer Vision and Pattern Recognition,
                  {CVPR} 2023, Vancouver, BC, Canada, June 17-24, 2023},
  pages        = {22563--22575},
  publisher    = {{IEEE}},
  year         = {2023},
  doi          = {10.1109/CVPR52729.2023.02161},
  bibsource    = {dblp computer science bibliography, https://dblp.org}
}

@inproceedings{phenaki,
  author       = {Ruben Villegas and
                  Mohammad Babaeizadeh and
                  Pieter{-}Jan Kindermans and
                  Hernan Moraldo and
                  Han Zhang and
                  Mohammad Taghi Saffar and
                  Santiago Castro and
                  Julius Kunze and
                  Dumitru Erhan},
  title        = {Phenaki: Variable Length Video Generation from Open Domain Textual
                  Descriptions},
  booktitle    = {The Eleventh International Conference on Learning Representations,
                  {ICLR} 2023, Kigali, Rwanda, May 1-5, 2023},
  publisher    = {OpenReview.net},
  year         = {2023},
  bibsource    = {dblp computer science bibliography, https://dblp.org}
}

@inproceedings{videocrafter,
  author       = {Haoxin Chen and
                  Yong Zhang and
                  Xiaodong Cun and
                  Menghan Xia and
                  Xintao Wang and
                  Chao Weng and
                  Ying Shan},
  title        = {VideoCrafter2: Overcoming Data Limitations for High-Quality Video
                  Diffusion Models},
  booktitle    = {{IEEE/CVF} Conference on Computer Vision and Pattern Recognition,
                  {CVPR} 2024, Seattle, WA, USA, June 16-22, 2024},
  pages        = {7310--7320},
  publisher    = {{IEEE}},
  year         = {2024},
  doi          = {10.1109/CVPR52733.2024.00698},
  bibsource    = {dblp computer science bibliography, https://dblp.org}
}

@inproceedings{text2videozero,
  author       = {Levon Khachatryan and
                  Andranik Movsisyan and
                  Vahram Tadevosyan and
                  Roberto Henschel and
                  Zhangyang Wang and
                  Shant Navasardyan and
                  Humphrey Shi},
  title        = {Text2Video-Zero: Text-to-Image Diffusion Models are Zero-Shot Video
                  Generators},
  booktitle    = {{IEEE/CVF} International Conference on Computer Vision, {ICCV} 2023,
                  Paris, France, October 1-6, 2023},
  pages        = {15908--15918},
  publisher    = {{IEEE}},
  year         = {2023},
  doi          = {10.1109/ICCV51070.2023.01462},
  bibsource    = {dblp computer science bibliography, https://dblp.org}
}

@article{wan22,
  author       = {Ang Wang and
                  Baole Ai and
                  Bin Wen and
                  Chaojie Mao and
                  Chen{-}Wei Xie et al.},
  title        = {Wan: Open and Advanced Large-Scale Video Generative Models},
  journal      = {CoRR},
  volume       = {abs/2503.20314},
  year         = {2025},
  url          = {https://doi.org/10.48550/arXiv.2503.20314},
  doi          = {10.48550/ARXIV.2503.20314},
  eprinttype    = {arXiv},
  eprint       = {2503.20314},
  bibsource    = {dblp computer science bibliography, https://dblp.org}
}

@article{veo,
  author       = {Thadd{\"{a}}us Wiedemer and
                  Yuxuan Li and
                  Paul Vicol and
                  Shixiang Shane Gu and
                  Nick Matarese and
                  Kevin Swersky and
                  Been Kim and
                  Priyank Jaini and
                  Robert Geirhos et al. },
  title        = {Video models are zero-shot learners and reasoners},
  journal      = {CoRR},
  volume       = {abs/2509.20328},
  year         = {2025},
  url          = {https://doi.org/10.48550/arXiv.2509.20328},
  doi          = {10.48550/ARXIV.2509.20328},
  eprinttype    = {arXiv},
  eprint       = {2509.20328},
  bibsource    = {dblp computer science bibliography, https://dblp.org}
}

@article{opensora2,
  author       = {Xiangyu Peng and
                  Zangwei Zheng and
                  Chenhui Shen and
                  Tom Young and
                  Xinying Guo et al. },
  title        = {Open-Sora 2.0: Training a Commercial-Level Video Generation Model
                  in {\textdollar}200k},
  journal      = {CoRR},
  volume       = {abs/2503.09642},
  year         = {2025},
  url          = {https://doi.org/10.48550/arXiv.2503.09642},
  doi          = {10.48550/ARXIV.2503.09642},
  eprinttype    = {arXiv},
  eprint       = {2503.09642},
  bibsource    = {dblp computer science bibliography, https://dblp.org}
}

@article{seedance,
  author       = {Team Seedance and
                  Heyi Chen and
                  Siyan Chen and
                  Xin Chen and
                  Yanfei Chen and
                  Ying Chen and
                  Zhuo Chen et al. },
  title        = {Seedance 1.5 pro: {A} Native Audio-Visual Joint Generation Foundation
                  Model},
  journal      = {CoRR},
  volume       = {abs/2512.13507},
  year         = {2025},
  url          = {https://doi.org/10.48550/arXiv.2512.13507},
  doi          = {10.48550/ARXIV.2512.13507},
  eprinttype    = {arXiv},
  eprint       = {2512.13507},
  bibsource    = {dblp computer science bibliography, https://dblp.org}
}

@article{kling,
  author       = {Kling Team},
  title        = {Kling-Omni Technical Report},
  journal      = {CoRR},
  volume       = {abs/2512.16776},
  year         = {2025},
  url          = {https://doi.org/10.48550/arXiv.2512.16776},
  doi          = {10.48550/ARXIV.2512.16776},
  eprinttype    = {arXiv},
  eprint       = {2512.16776},
  bibsource    = {dblp computer science bibliography, https://dblp.org}
}

@misc{T2V-opt,
      title={T2V-OptJail: Discrete Prompt Optimization for Text-to-Video Jailbreak Attacks}, 
      author={Jiayang Liu and Siyuan Liang and Shiqian Zhao and Rongcheng Tu and Wenbo Zhou and Aishan Liu and Dacheng Tao and Siew Kei Lam},
      year={2025},
      eprint={2505.06679},
      archivePrefix={arXiv},
      primaryClass={cs.CV},
      url={https://arxiv.org/abs/2505.06679}, 
}

@article{runwayevil,
  author       = {Songping Wang and
                  Rufan Qian and
                  Yueming Lyu and
                  Qinglong Liu and
                  Linzhuang Zou and
                  Jie Qin and
                  Songhua Liu and
                  Caifeng Shan},
  title        = {RunawayEvil: Jailbreaking the Image-to-Video Generative Models},
  journal      = {CoRR},
  volume       = {abs/2512.06674},
  year         = {2025},
  url          = {https://doi.org/10.48550/arXiv.2512.06674},
  doi          = {10.48550/ARXIV.2512.06674},
  eprinttype    = {arXiv},
  eprint       = {2512.06674},
  bibsource    = {dblp computer science bibliography, https://dblp.org}
}

@inproceedings{t2vsafetybench,
  author       = {Yibo Miao and
                  Yifan Zhu and
                  Lijia Yu and
                  Jun Zhu and
                  Xiao{-}Shan Gao and
                  Yinpeng Dong},
  editor       = {Amir Globersons and
                  Lester Mackey and
                  Danielle Belgrave and
                  Angela Fan and
                  Ulrich Paquet and
                  Jakub M. Tomczak and
                  Cheng Zhang},
  title        = {T2VSafetyBench: Evaluating the Safety of Text-to-Video Generative
                  Models},
  booktitle    = {Advances in Neural Information Processing Systems 38: Annual Conference
                  on Neural Information Processing Systems 2024, NeurIPS 2024, Vancouver,
                  BC, Canada, December 10 - 15, 2024},
  year         = {2024},
  bibsource    = {dblp computer science bibliography, https://dblp.org}
}

@inproceedings{unsafebench,
  author       = {Yiting Qu and
                  Xinyue Shen and
                  Yixin Wu and
                  Michael Backes and
                  Savvas Zannettou and
                  Yang Zhang},
  editor       = {Chun{-}Ying Huang and
                  Jyh{-}Cheng Chen and
                  Shiuh{-}Pyng Shieh and
                  David Lie and
                  V{\'{e}}ronique Cortier},
  title        = {UnsafeBench: Benchmarking Image Safety Classifiers on Real-World and
                  AI-Generated Images},
  booktitle    = {Proceedings of the 2025 {ACM} {SIGSAC} Conference on Computer and
                  Communications Security, {CCS} 2025, Taipei, Taiwan, October 13-17,
                  2025},
  pages        = {3221--3235},
  publisher    = {{ACM}},
  year         = {2025},
  doi          = {10.1145/3719027.3765088},
  bibsource    = {dblp computer science bibliography, https://dblp.org}
}

@inproceedings{vidprom,
  author       = {Wenhao Wang and
                  Yi Yang},
  editor       = {Amir Globersons and
                  Lester Mackey and
                  Danielle Belgrave and
                  Angela Fan and
                  Ulrich Paquet and
                  Jakub M. Tomczak and
                  Cheng Zhang},
  title        = {VidProM: {A} Million-scale Real Prompt-Gallery Dataset for Text-to-Video
                  Diffusion Models},
  booktitle    = {Advances in Neural Information Processing Systems 38: Annual Conference
                  on Neural Information Processing Systems 2024, NeurIPS 2024, Vancouver,
                  BC, Canada, December 10 - 15, 2024},
  year         = {2024},
  bibsource    = {dblp computer science bibliography, https://dblp.org}
}

@article{scenespilt,
  author       = {Wonjun Lee and
                  Haon Park and
                  Doehyeon Lee and
                  Bumsub Ham and
                  Suhyun Kim},
  title        = {Jailbreaking on Text-to-Video Models via Scene Splitting Strategy},
  journal      = {CoRR},
  volume       = {abs/2509.22292},
  year         = {2025},
  url          = {https://doi.org/10.48550/arXiv.2509.22292},
  doi          = {10.48550/ARXIV.2509.22292},
  eprinttype    = {arXiv},
  eprint       = {2509.22292},
  bibsource    = {dblp computer science bibliography, https://dblp.org}
}

@inproceedings{SurrogatePrompt,
  author       = {Zhongjie Ba and
                  Jieming Zhong and
                  Jiachen Lei and
                  Peng Cheng and
                  Qinglong Wang and
                  Zhan Qin and
                  Zhibo Wang and
                  Kui Ren},
  editor       = {Bo Luo and
                  Xiaojing Liao and
                  Jun Xu and
                  Engin Kirda and
                  David Lie},
  title        = {SurrogatePrompt: Bypassing the Safety Filter of Text-to-Image Models
                  via Substitution},
  booktitle    = {Proceedings of the 2024 on {ACM} {SIGSAC} Conference on Computer and
                  Communications Security, {CCS} 2024, Salt Lake City, UT, USA, October
                  14-18, 2024},
  pages        = {1166--1180},
  publisher    = {{ACM}},
  year         = {2024},
  url          = {https://doi.org/10.1145/3658644.3690346},
  doi          = {10.1145/3658644.3690346},
  bibsource    = {dblp computer science bibliography, https://dblp.org}
}

@article{Divide-and-Conquer,
  author       = {Yimo Deng and
                  Huangxun Chen},
  title        = {Divide-and-Conquer Attack: Harnessing the Power of {LLM} to Bypass
                  the Censorship of Text-to-Image Generation Model},
  journal      = {CoRR},
  volume       = {abs/2312.07130},
  year         = {2023},
  url          = {https://doi.org/10.48550/arXiv.2312.07130},
  doi          = {10.48550/ARXIV.2312.07130},
  eprinttype    = {arXiv},
  eprint       = {2312.07130},
  bibsource    = {dblp computer science bibliography, https://dblp.org}
}

@inproceedings{SneakyPrompt,
  author       = {Yuchen Yang and
                  Bo Hui and
                  Haolin Yuan and
                  Neil Gong and
                  Yinzhi Cao},
  title        = {SneakyPrompt: Jailbreaking Text-to-image Generative Models},
  booktitle    = {{IEEE} Symposium on Security and Privacy, {SP} 2024, San Francisco,
                  CA, USA, May 19-23, 2024},
  pages        = {897--912},
  publisher    = {{IEEE}},
  year         = {2024}
}

@inproceedings{Ring-a-bell,
  author       = {Yu{-}Lin Tsai and
                  Chia{-}Yi Hsu and
                  Chulin Xie and
                  Chih{-}Hsun Lin and
                  Jia{-}You Chen and
                  Bo Li and
                  Pin{-}Yu Chen and
                  Chia{-}Mu Yu and
                  Chun{-}Ying Huang},
  title        = {Ring-A-Bell! How Reliable are Concept Removal Methods For Diffusion
                  Models?},
  booktitle    = {The Twelfth International Conference on Learning Representations,
                  {ICLR} 2024, Vienna, Austria, May 7-11, 2024},
  publisher    = {OpenReview.net},
  year         = {2024},
}

@misc{gpt5card,
      title={GPT-5}, 
      author={Aaditya Singh and Adam Fry and Adam Perelman and Adam Tart et al.},
      year={2025},
      eprint={2601.03267},
      archivePrefix={arXiv},
      primaryClass={cs.CL},
      url={https://arxiv.org/abs/2601.03267}, 
}

@misc{blacklist,
  author = {Rojit George.},
  title = {Nsfw words list on GitHub.},
  year = {2020},
  howpublished = {\url{https://github.com/rrgeorge-pdcontributions/NSFW-Words-List/blob/master/nsfw_list.txt}},
}

@misc{nsfw-classifier,
      title={DiffGuard: Text-Based Safety Checker for Diffusion Models}, 
      author={Massine El Khader and Elias Al Bouzidi and Abdellah Oumida and Mohammed Sbaihi and Eliott Binard and Jean-Philippe Poli and Wassila Ouerdane and Boussad Addad and Katarzyna Kapusta},
      year={2025},
      eprint={2412.00064},
      archivePrefix={arXiv},
      primaryClass={cs.CV},
      url={https://arxiv.org/abs/2412.00064}, 
}

@misc{image-classifier,
  author = {Lakshay Chhabra.},
  title = {Nsfw image classifier on GitHub.},
  year = {2020},
  howpublished = {\url{https://github.com/lakshaychhabra/NSFW-Detection-DL}},
}

@misc{llamaguard38b,
  title =         {The Llama 3 Herd of Models},
  author =        {Llama Team, AI @ Meta},
  year =          {2024},
  eprint =        {2407.21783},
  archivePrefix = {arXiv},
  primaryClass =  {cs.AI},
  url =           {https://arxiv.org/abs/2407.21783}
}

@misc{spark,
      title={SPARK: Jailbreaking T2V Models by Synergistically Prompting Auditory and Recontextualized Knowledge}, 
      author={Zonghao Ying and Moyang Chen and Nizhang Li and Zhiqiang Wang and Wenxin Zhang and Quanchen Zou and Zonglei Jing and Aishan Liu and Xianglong Liu},
      year={2026},
      eprint={2511.13127},
      archivePrefix={arXiv},
      primaryClass={cs.CV},
      url={https://arxiv.org/abs/2511.13127}, 
}

@inproceedings{sql-injection,
    title = "{SQL} Injection Jailbreak: A Structural Disaster of Large Language Models",
    author = "Zhao, Jiawei  and
      Chen, Kejiang  and
      Zhang, Weiming  and
      Yu, Nenghai",
    editor = "Che, Wanxiang  and
      Nabende, Joyce  and
      Shutova, Ekaterina  and
      Pilehvar, Mohammad Taher",
    booktitle = "Findings of the Association for Computational Linguistics: ACL 2025",
    month = jul,
    year = "2025",
    address = "Vienna, Austria",
    publisher = "Association for Computational Linguistics",
    url = "https://aclanthology.org/2025.findings-acl.358/",
    doi = "10.18653/v1/2025.findings-acl.358",
    pages = "6871--6891",
    ISBN = "979-8-89176-256-5",
}

@article{gcg,
  author       = {Andy Zou and
                  Zifan Wang and
                  J. Zico Kolter and
                  Matt Fredrikson},
  title        = {Universal and Transferable Adversarial Attacks on Aligned Language
                  Models},
  journal      = {CoRR},
  volume       = {abs/2307.15043},
  year         = {2023},
  url          = {https://doi.org/10.48550/arXiv.2307.15043},
  doi          = {10.48550/ARXIV.2307.15043},
  eprinttype    = {arXiv},
  eprint       = {2307.15043},
}

@article{T2Vshield,
  author       = {Siyuan Liang and
                  Jiayang Liu and
                  Jiecheng Zhai and
                  Tianmeng Fang and
                  Rongcheng Tu and
                  Aishan Liu and
                  Xiaochun Cao and
                  Dacheng Tao},
  title        = {T2VShield: Model-Agnostic Jailbreak Defense for Text-to-Video Models},
  journal      = {CoRR},
  volume       = {abs/2504.15512},
  year         = {2025},
  url          = {https://doi.org/10.48550/arXiv.2504.15512},
  doi          = {10.48550/ARXIV.2504.15512},
  eprinttype    = {arXiv},
  eprint       = {2504.15512},
}

@misc{videollama2,
      title={VideoLLaMA 2: Advancing Spatial-Temporal Modeling and Audio Understanding in Video-LLMs}, 
      author={Zesen Cheng and Sicong Leng and Hang Zhang and Yifei Xin and Xin Li and Guanzheng Chen and Yongxin Zhu and Wenqi Zhang and Ziyang Luo and Deli Zhao and Lidong Bing},
      year={2024},
      eprint={2406.07476},
      archivePrefix={arXiv},
      primaryClass={cs.CV},
      url={https://arxiv.org/abs/2406.07476}, 
}

\end{document}